\documentclass[12pt]{article}
\usepackage{hyperref}
\usepackage{epsfig}
\usepackage{float}
\usepackage{empheq}
\usepackage{bbold}

\usepackage[utf8]{inputenc}
\usepackage{amsmath}

\usepackage{caption}

\usepackage{amsmath}
\usepackage{amssymb}
\usepackage{graphicx}
\newlength{\abstractwidth}
\renewcommand{\thefootnote}{\fnsymbol{footnote}}
\renewcommand{\thanks}[1]{\footnote{#1}} 
\newcommand{\starttext}{
\setcounter{footnote}{0}
\renewcommand{\thefootnote}{\arabic{footnote}}}

\newcommand{\be}{\begin{equation}}
\newcommand{\bea}{\begin{eqnarray}}
\newcommand{\eea}{\end{eqnarray}}
\newcommand{\beq}{\begin{equation}}
\newcommand{\ee}{\end{equation}}

\newcommand*\widefbox[1]{\fbox{\hspace{2em}#1\hspace{2em}}}

\def\eq{&=&}

\def\simleq{\; \raise0.3ex\hbox{$<$\kern-0.75em
\raise-1.1ex\hbox{$\sim$}}\; }
\def\simgeq{\; \raise0.3ex\hbox{$>$\kern-0.75em
\raise-1.1ex\hbox{$\sim$}}\; }

\def\bi{\begin{itemize}}
\def\ei{\end{itemize}}
\def\S{Schwarzschild}

\def\CR{{\cal{R}}}

\def\Tr{\rm Tr \it}
\def\bsub{ \begin{subequations}
\begin{empheq}[box=\widefbox]{align}  }
\def\esub{ \end{empheq}
\end{subequations}}

\def\1{\(  \mathbb{1} \)}

 \def\lf{\left(}
    \def\rg{\right)}

  \def\bn{\bigskip \noindent}

 \def\bm{\begin{bmatrix}}
 \def\em{\end{bmatrix}}

\makeatletter
\g@addto@macro\normalsize{%
  \setlength\abovedisplayskip{10pt}
  \setlength\belowdisplayskip{20pt}
  \setlength\abovedisplayshortskip{10pt}
  \setlength\belowdisplayshortskip{20pt}
}
\makeatother

\usepackage{color}

\begin{document}


  
\begin{titlepage}

\rightline{}
\bigskip
\bigskip\bigskip\bigskip\bigskip
\bigskip

\centerline{\Large \bf {Black Holes Hint Towards De Sitter-Matrix Theory }}

\bn

\bigskip
\begin{center}
\bf      Leonard Susskind  \rm

\bigskip
Stanford Institute for Theoretical Physics and Department of Physics, \\
Stanford University,
Stanford, CA 94305-4060, USA \\

and

Google, Mountain View, CA

\end{center}

\bn

\begin{abstract}

De Sitter black holes and other non-perturbative configurations can be used to probe the holographic degrees of freedom  of de Sitter space.   For small black holes  evidence was first given in seminal  work of Banks, Fiol, and Morrise; and  followups by Banks and Fischler; showing  that dS is described by a form of matrix theory. For large black holes the evidence given here is new: Gravitational calculations and matrix theory calculations of the rates of exponentially  rare fluctuations match one another in surprising detail.
The occurrence of the  Nariai geometry and the ``inside-out" transition are especially interesting examples which I explain.

\end{abstract}

\end{titlepage}

\starttext \baselineskip=17.63pt \setcounter{footnote}{0}

\tableofcontents

\Large


\section{Entanglement in de Sitter Space}

In this paper I will assume that there is  a holographic description of the static patches of  four-dimensional  de Sitter space\footnote{The various mechanisms and calculations described in this paper apply to four dimensions. Generalization to other number of dimensions is non-trivial and I will not undertake the task here. }; but unlike AdS, de Sitter space has no asymptotic  boundary where the degrees of freedom are located. Instead, the holographic degrees of freedom are nominally located on the boundary of the static patch (SP) (see for example \cite{Dyson:2002pf}\cite{Banks:2006rx}\cite{Susskind:2011ap}\cite{Banks:2016taq}\cite{Banks:2020zcr}\cite{Susskind:2021omt}); that is to say, the stretched horizon. 

\begin{figure}[H]
\begin{center}
\includegraphics[scale=.28]{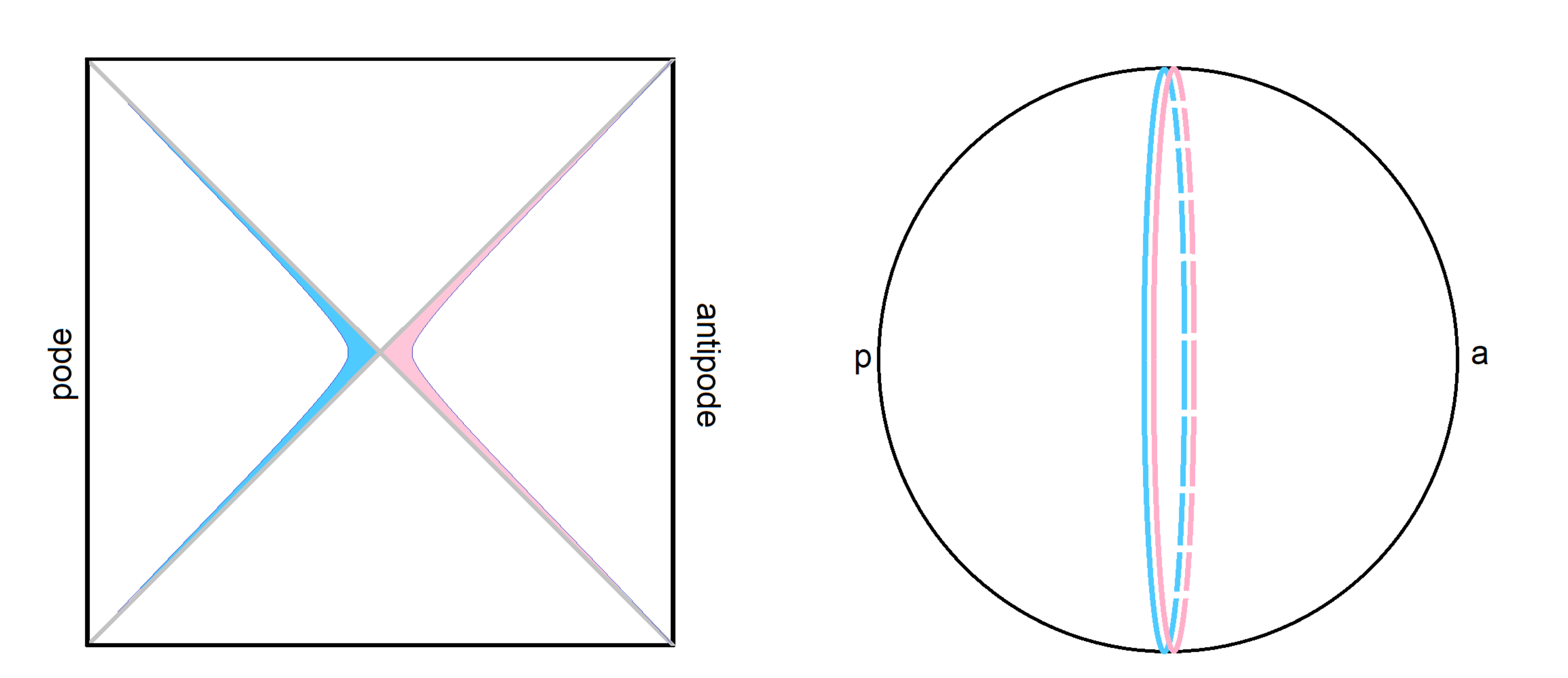}
\caption{The left panel shows the Penrose diagram for de Sitter space with a particular choice of opposing static patches. The blue and pink regions are the stretched horizons of the two SPs.  The right panel shows the spatial geometry of a time-symmetric slice. The blue and pink surfaces represent the stretched horizons.}
\label{basic}
\end{center}
\end{figure}

Static patches come in opposing pairs. To account for the pair,  two sets of degrees of freedom are required. The Penrose diagram of de Sitter space in figure \ref{basic} shows such a pair of SPs along with their stretched horizons. The center of the SPs (sometimes thought of as the points where  observers are located) will be called the pode and the antipode.

Although it is clear from the Penrose diagram that the two SPs are entangled in the thermofield-double state, no clear framework similar to the Ryu-Takyanagi formula has been formulated for de Sitter space. This paper is not primarily about such a de Sitter generalization of the RT framework but I will briefly sketch what such a generalization looks like.

 We assume  that the entanglement entropy of the two sides---pode and antipode---is proportional to the minimum area of a surface homologous to the boundary of one of the two components---let us say the pode side. But what do we mean by the boundary? The full  spatial slice at $t=0$  has no boundary, but the static patch is bounded by the blue stretched horizon. Thus we try the following formulation:

\bn 
\it The entanglement entropy of the pode-antipode systems is $1/4G$ times the minimal area
of a surface homologous to the   stretched horizon (of either side).
\rm

\begin{figure}[H]
\begin{center}
\includegraphics[scale=.4]{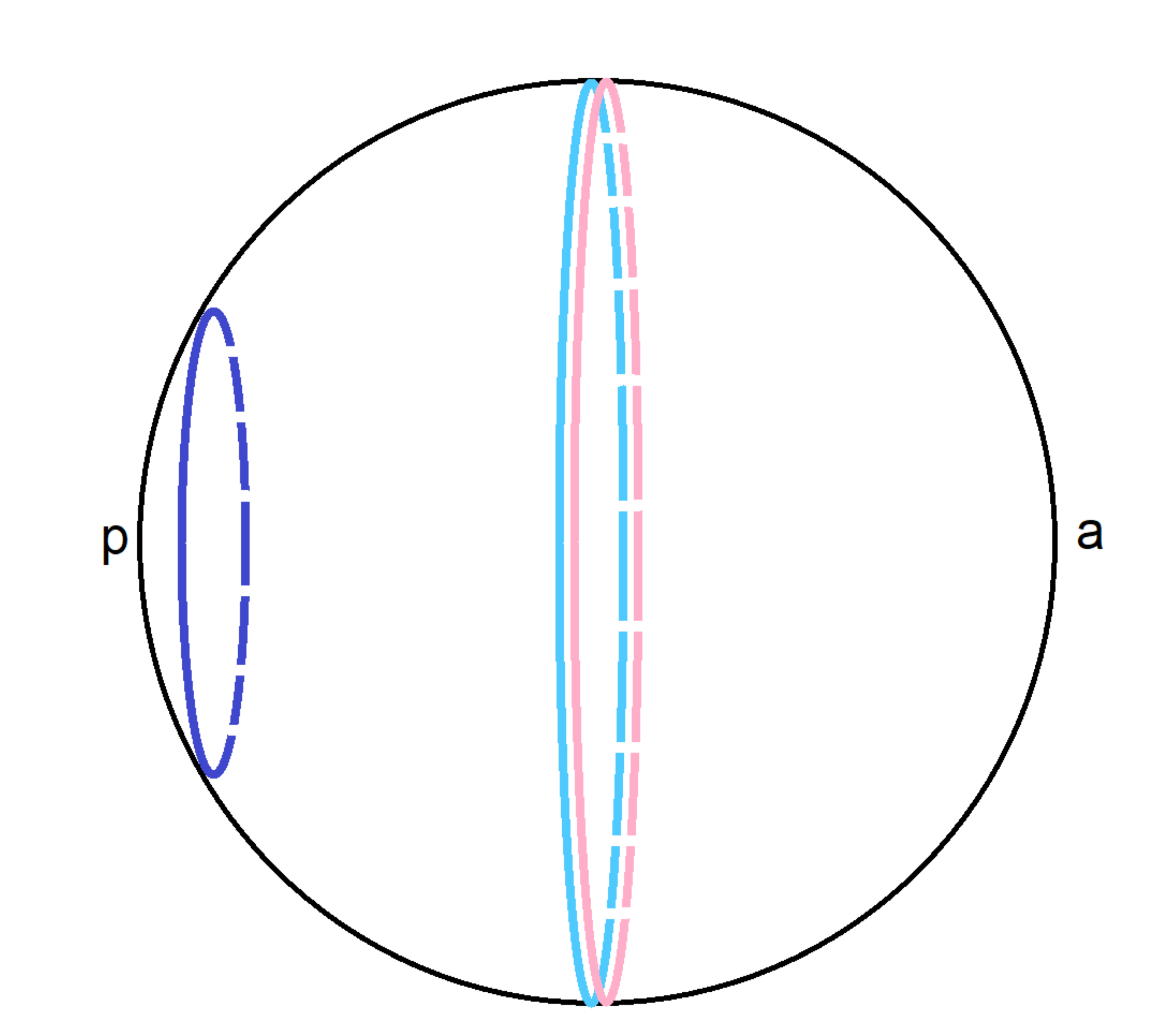}
\caption{A $t=0$ slice of dS and the stretched horizons shown as light blue and pink great circles. The dark blue surface is homologous to the light blue horizon. It  can be shrunk to a point.}
\label{homologous1}
\end{center}
\end{figure}

\bn
This however will not work. Figure \ref{homologous1} shows the spatial slice and the adjacent pair of stretched horizons. The dark blue curve represents a surface homologous to the pode's stretched horizon. It is obvious that that curve can be shrunk to zero, which if the above formulation were correct would imply vanishing entanglement between the pode and antipode static patches.   
\rm

\bn

To do better first separate the two stretched horizons a bit. This is a natural thing to do  since they will   separate after a short time, as is obvious from figure \ref{basic}. Let us now reformulate a dS-improved version of the RT principle:

\bn 
\it The entanglement entropy of the pode-antipode systems is $1/4G$ times the minimal area
of a surface homologous to the stretched horizon of the pode, and lying between the two sets of degrees of freedom, i.e., between the two stretched horizons. 
\rm

\bn

\begin{figure}[H]
\begin{center}
\includegraphics[scale=.3]{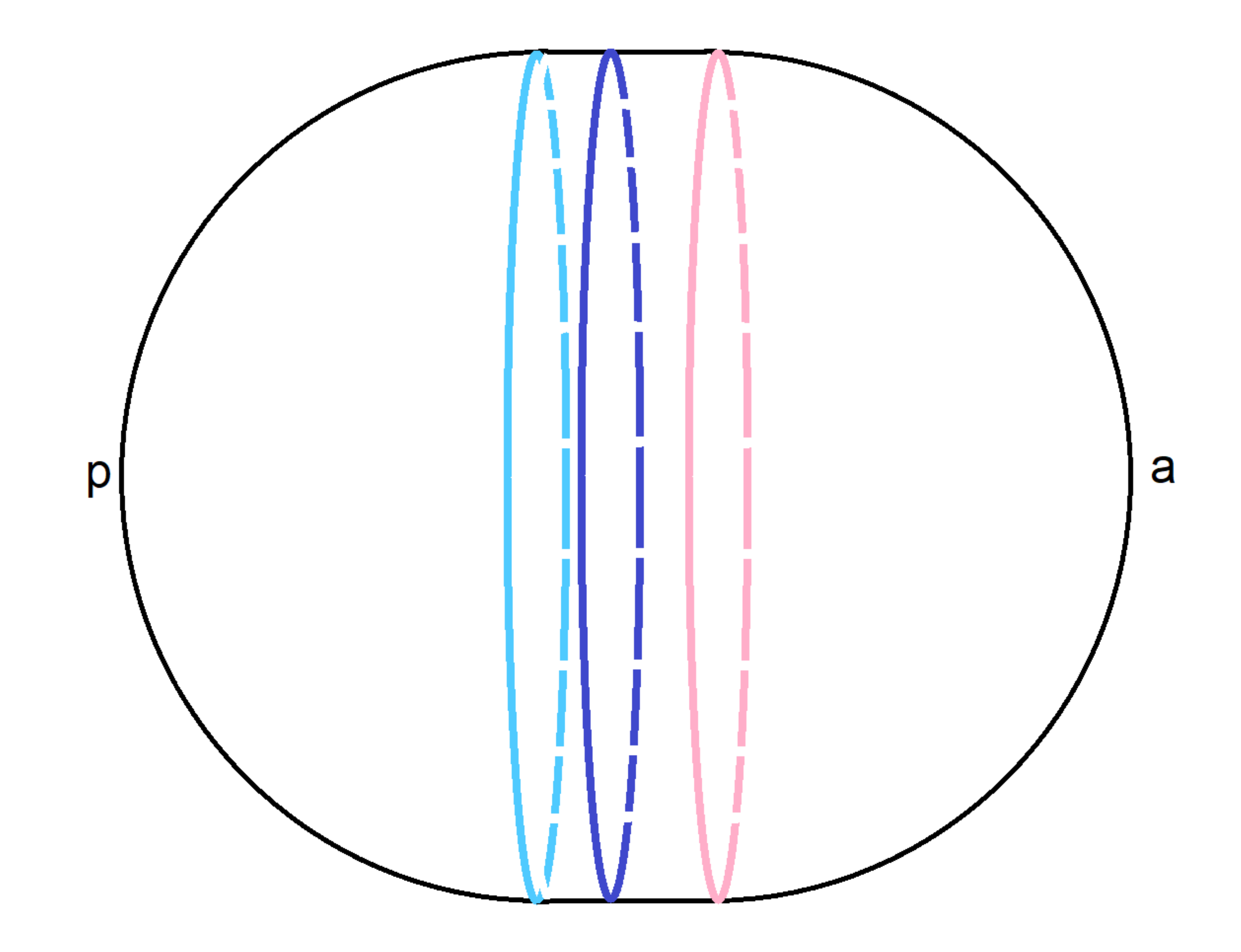}
\caption{The dark blue curve represents the minimal  surface lying between the two stretched horizons shown in light blue and pink. }
\label{homologous3}
\end{center}
\end{figure}

This version of the RT principle is illustrated in figure \ref{homologous3}

\bn
It is evident from the figure that the area of the dSRT surface is the area of the horizon.
This gives the entanglement entropy that we expect \cite{Gibbons:1977mu}, namely $$\frac{\rm Horizon \ Area}{4G}.$$
One thing to note, is that in anti-de Sitter space the phrase ``lying between the two sets of degrees of freedom" is redundant. The  degrees of freedom lie at the asymptotic boundary and any minimal surface will necessarily  lie between them.

\begin{figure}[H]
\begin{center}
\includegraphics[scale=.4]{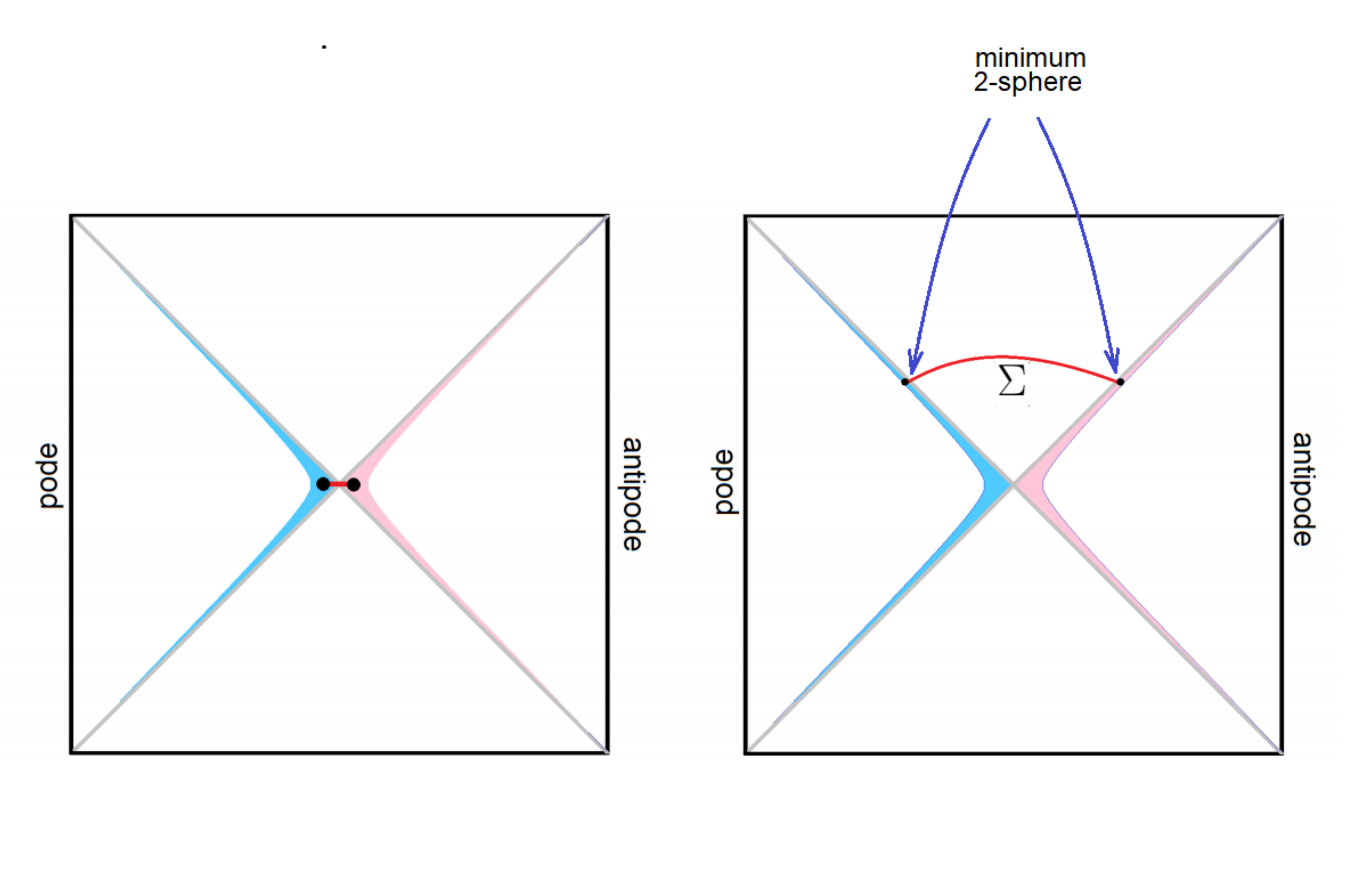}
\caption{ In both panels the black dots represent the anchoring points of a space-like surface $\Sigma$ connecting the two horizons at a particular time. The minimal two-sphere cutting $\Sigma$ lies at the anchoring points.}
\label{RTdS}
\end{center}
\end{figure}

This version of the de Sitter RT formula is sufficient for time-independent geometries. A more general ``maxmin"  formulation  goes as follows:
Pick a time on the stretched horizons and anchor a three-dimensional surface $\Sigma$ connecting  the two. This is shown in figure \ref{RTdS}. 
Find the minimum-area two-dimensional sphere that cuts the three-dimensional surface $\Sigma$ and call its area $A_{min}(\Sigma).$ It is not hard to show that  the minimum area sphere hugs one of the two horizons as in figure \ref{RTdS}. The reason is that in de Sitter space the local 2-sphere grows (exponentially)  as one moves behind the horizon.

Now maximize $A_{min}(\Sigma)$ over all space-like $\Sigma.$ Call the resulting area $$A_{maxmin}.$$  The entanglement entropy between the pode and antipode static patches is,
\be 
S_{ent} = \frac{A_{maxmin}}{4G}.
\ee
Because $A_{min}(\Sigma)$ occurs at the anchoring points  the maximization of $A_{min}(\Sigma)$ is redundant in the case depicted in figure \ref{basic}.

It should be possible to generalize the dSRT formula to include bulk entanglement term, but I will save this for another time.
\bn

Now we turn to the main subject of this paper---dS black holes and their implications for dS holography.

\section{From Small Black Holes to Nariai}
The properties of black holes in four-dimensional de Sitter space provide hints about  the holographic  degrees of freedom and their dynamics.  These hints will lead us to a remarkably general conclusion: the underlying holographic description of de Sitter space must be a form a matrix quantum mechanics.

The \S \ de Sitter metric is given by,
\bea
ds^2  &=& -f(r)dt^2 +f(r)^{-1}dr^2 +r^2 d\Omega^2 \cr \cr
f(r) \eq 1-\frac{r^2}{R^2} -\frac{2MG}{r}
\label{SdS}
\eea
where $R$ is the de Sitter radius, $M$ is the black hole mass, and $G$ is Newton's constant.

\begin{figure}[H]
\begin{center}
\includegraphics[scale=.4]{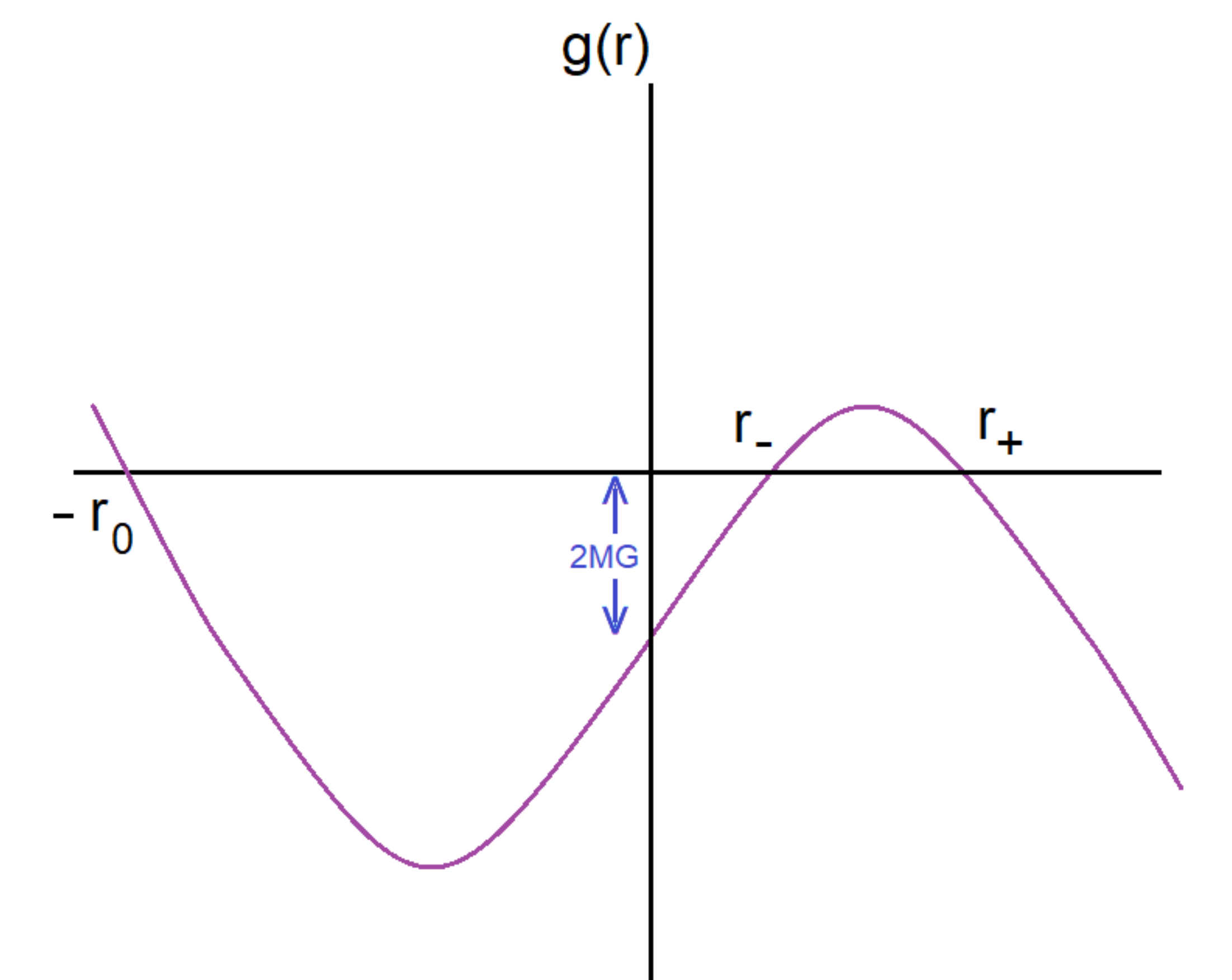}
\caption{The function $g(r)$ and its zeros, $r_0$ and $r_{\pm}.$}
\label{cubic}
\end{center}
\end{figure}

There are two horizons, the larger  cosmic horizon and the smaller  black hole horizon. The horizons are defined by $f(r) =0.$ Defining $g(r) = rf(r)$ the horizon condition becomes,
\be  
g(r) = r-r^3/R^2-2MG =0.
\label{gee}
\ee
The function $g(r)$ is shown in figure \ref{cubic}.

\bn
For values of $M$ satisfying 
\be
0<\frac{MG}{R} < \frac{1}{3\sqrt{3}},
\label{M range}
\ee  
equation \eqref{gee} has three solutions, two with  positive values of $r,$ and one with negative $r.$  The two positive solutions, $r_-$ and $r_+$ define the black hole horizon and the cosmic horizon respectively. The negative  solution, $r=-r_0,$ is unphysical. Outside the range \eqref{M range} the metric has a naked singularity. 
Given the values of $R$ (or equivalently the cosmological constant) and $G,$ there is only one parameter in the metric, namely $M.$  Alternatively we may choose the independent parameter to be either $r_+,$  $r_-,$ or  the dimensionless  parameter $x$ 
defined by,
\be 
x  \equiv (r_+ - r_-)/R.
\label{x}
\ee

The variable $x$ runs from   $x=-1$ to $x=+1.$ 
Over this range the mass $M$ runs over its allowable values \eqref{M range} twice: once for $x<0$ and once for $x>0.$ As $x$ increases from $-1$ to $0$ the black hole horizon $r_-$ grows, and the cosmic horizon $r_+$ shrinks, so that the two become equal at $x=0.$
When $x$ becomes positive the two horizons are exchanged so that $r_- > r_+$. Beyond that   $r_+$ becomes the black hole horizon,  $r_-$  the cosmic horizon.

There are two possible ways to think about this. In the first we assume that the range $x>0$ is redundant and simply describes the same states that were covered for $x<0$; roughly speaking we think of the choice of the sign of $x$ as a gauge choice.
The second  possibility is that the two ranges are physically different configurations. We will adopt the latter viewpoint in this paper.

The cubic function $g(r)$ in \eqref{gee} may be written as a product,
\be 
g(r) =-\frac{(r-r_+)(r-r_-)(r+r_0)  }{R^2.}  
\label{gee product}
\ee
Matching \eqref{gee} with \eqref{gee product} we find the following relations,
\bea 
r_0 \eq r_+ + r_-  \ \ \ \ \ \ \ \ \ \ \ \ \ \  \ \ \ (a)  \cr \cr
R^2 \eq r_+^2 +r_-^2 +r_+r_-  \ \ \ \ \ \    (b)  \cr \cr 
2MG R^2 \eq r_+ r_- (r_+ + r_-)  \ \ \ \ \ \ \ \  (c) \cr \cr
 x^2 &=&  \frac{r_+^2 + r_-^2 -2 r_+ r_-}{R^2}      \ \ \ \ \ (d)
\label{match}
\eea
The last of these equations---\eqref{match}$(d)$---is just the square of the defining relation \eqref{x}. By combining \eqref{match}$(d)$ and \eqref{match}$(b)$ we  find the relation,
\be 
R^2 -r_+^2 - r_-^2 =     \frac{ R^2}{3}\lf 1 -x^2 \rg
\label{r and x}
\ee

The significance of this equation will become clear in subsection \ref{large}.

\section{Entropy Deficit}
In thermal equilibrium the entropy is maximized subject to the constraint of a given average energy. In the context of the static patch, the equilibrium entropy is the usual de Sitter entropy which I will call $S_0,$ 
\be 
S_0 =\frac{ \rm{area}}{4G} =\frac{\pi R^2}{G}.
\label{S=pRR/G}
\ee
Fluctuations may occur in which the entropy is decreased to a smaller value $S_1.$ The probability for such a fluctuation is given in terms of the entropy deficit, 

\be 
\Delta S \equiv S_0 -S_1
\label{entdeft}
\ee
by
\be 
{\rm Probability} = e^{-\Delta S}
\label{P=e-DS}
\ee

\subsection{Small Black Holes}
By a small black hole I mean one whose mass in Planck units  is fixed as $R$ becomes large. It may also be defined by its entropy being parametrically order $1$ as the de Sitter entropy is taken to infinity.

Let us consider the probability for a fluctuation in which a black hole of mass $M$  appears at the pode\footnote{The center of the causal patch at $r=0$. The center of the opposing static patch is called the antipode  \cite{Susskind:2021omt}.}. The thermal equilibrium entropy of de Sitter space is given in \eqref{S=pRR/G}.
To compute the entropy $S_1$ of a state with a black hole at the pode we use equation \eqref{gee} to compute $r_+.$ To lowest order in $M$ we find,
\be 
r_+ = R-MG.
\label{r+=R-MG}
\ee
The area and entropy of the cosmic horizon to lowest order are,
\bea 
\rm{area} \eq  4\pi (R^2 - 2RMG) \cr \cr
S \eq  \frac{\pi (R^2 - 2RMG)}{G},
\label{A=4p(R2-2MG)}
\eea
and the entropy deficit is,
\be 
 \Delta S   =2\pi R M.   
\label{s=2Prm}
\ee
The probability for the fluctuation is 
\be 
{\rm{Prob}} = e^{-\Delta S} = e^{-2\pi RM},
\label{P=e-2pRM}
\ee
which is also the Boltzmann weight $e^{-\beta M}.$

Another interesting form for the entropy deficit for small black holes can easily be derived and is given by,
\be 
\Delta S = \sqrt{Ss},
\label{DS=sqrtSs}
\ee
where $s$ is the black hole entropy $4\pi M^2 G.$

Note that for fluctuations involving  small black holes of fixed mass, the entropy deficit goes as $S^{1/2}$. 
For example the probability of a Planck mass black hole ($s=1$) is\footnote{In this paper the notation $\exp$ is synonymous with ``exponential in." Thus $e^S,$ $e^{S/3}$, $e^{2S}$ are all $\exp{S}.$},
\be 
  {\rm{Prob}} \sim \exp{(-\sqrt{S})}. 
  \label{DSeqsqrts}
\ee

\bn  

Equation \eqref{DS=sqrtSs} is a strong  hint about the holographic degrees of freedom of de Sitter space. 
It is quite unusual in the way it depends on the entropies of not only the black hole, but also on the entropy of the cosmic horizon. 
The question it raises is: without direct reference to gravity, what kind of holographic degrees of freedom can lead to such a  relation?  
 
 Following \cite{Banks:1996vh}, we will see in 
 section \ref{dSmatrix} that the answer is matrix degrees of freedom 
 similar to those of BFSS M(atrix) theory. Moreover \ref{DSeqsqrts} is suggestive of instantons in large-$N$ matrix quantum mechanics and  large-$N$ gauge theories.  In  subsection \ref{large} we will see another more detailed  relation for the entropy deficit of large black holes that even more strongly supports the claim for matrix degrees of freedom.

\subsection{Large Black Holes} \label{large}

By a  large  black hole I mean one with \S \ radius parametrically of order the de Sitter radius $R.$ Equivalently, the entropy of a large black hole is a fixed fraction of the de Sitter entropy.
Such black holes are characterized  by fixed values of the dimensionless variable $x$ defined by \eqref{x}. At $x  \approx-1 $ the black hole  radius is very small compared  to the radius of the cosmic horizon.  At $x=0$ the two horizon areas become equal, and of order $R^2$.  At that point the geometry is called the Nariai geometry. Its properties are reviewed in the appendix. 

For $x>0$ the black hole and cosmic horizon switch roles.  As mentioned earlier we will consider $x=-1$ and $x=1$ to be different states.

The entropy of a horizon of radius $r$ is given by,
\be 
S= { \frac{\rm area \it  }{4G}} = \pi r^2.
\label{S(r)}
\ee
Applying this to the original de Sitter horizon with radius $R,$ and to the horizons $r_{\pm}$ of the \S -dS geometry,
\bea
S_0\eq    \frac{\pi R^2}{G}       \cr \cr
S_+ \eq    \frac{\pi r_+^2}{G}     \cr \cr
S_- \eq      \frac{\pi r_-^2}{G}   .
\label{S0+-}
\eea

The total entropy is the sum of the black hole and cosmic entropies,
\be 
S(x)= S_+(x) + S_-(x).
\label{SeqSpm}
\ee

Armed with these relations we may write equation \eqref{r and x}  in the surprisingly simple form,
\bea
\Delta S(x) 
&=&  \pi R^2  \lf  \frac{1 -x^2}{3G} \rg \cr \cr
\eq \frac{S_0}{3}(1-x^2)
\label{S(1-xx)}
\eea

\begin{figure}[H]
\begin{center}
\includegraphics[scale=.4]{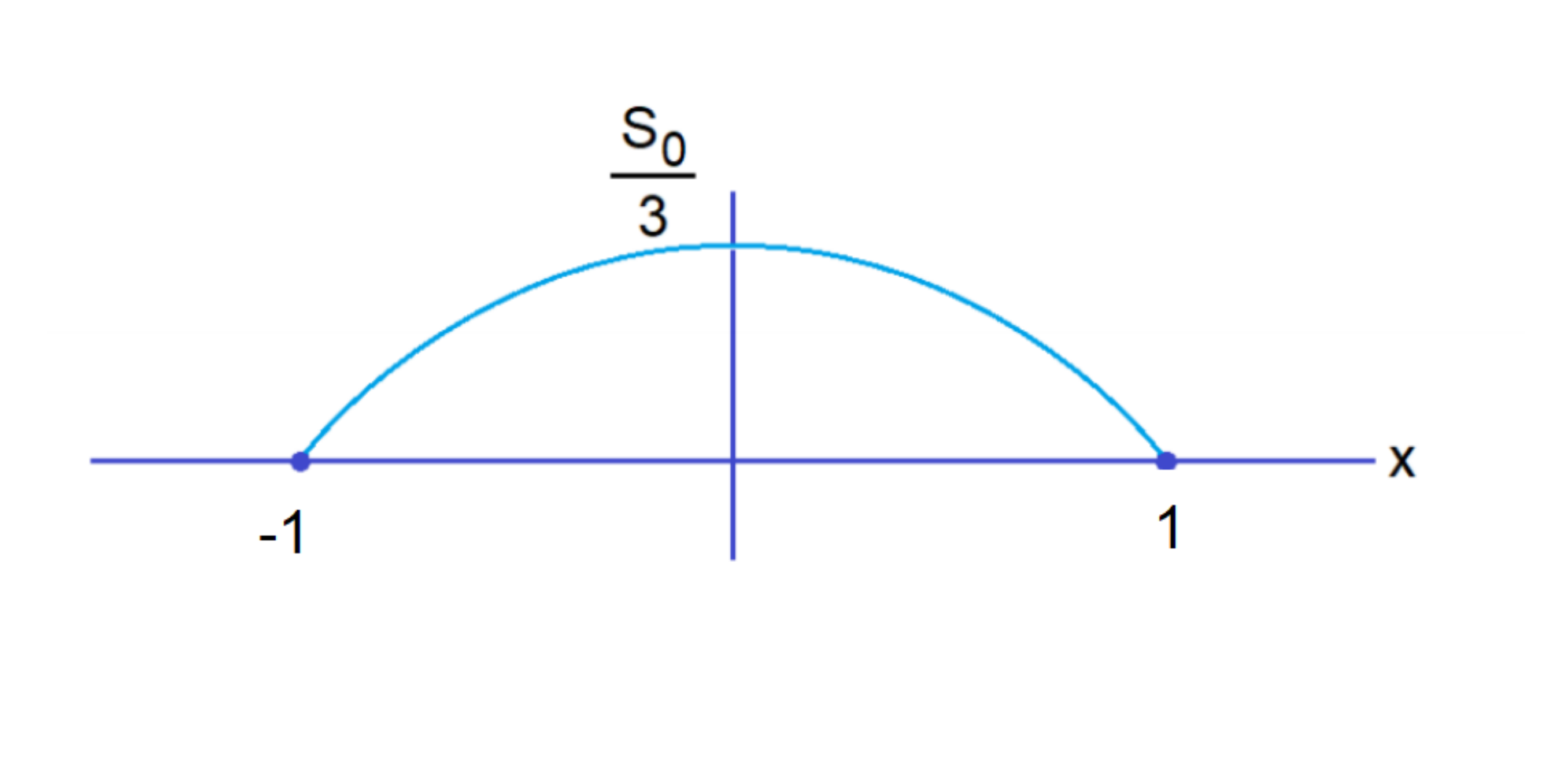}
\caption{The entropy deficit as a function of $x.$}
\label{D(x)}h
\end{center}
\end{figure}

Equation \eqref{S(1-xx)}  is illustrated
in figure \ref{D(x)}.

\bn
Equation \eqref{S(1-xx)} may also be written,
\be 
\Delta S(x) = \Delta S_N \ (1-x^2),
\label{GRcurve}
\ee
where $\Delta S_N = \frac{1}{3}S_0$ is  the entropy deficit of the  Nariai geometry.
 Note that $\Delta S$  is symmetric under $x \to -x$ and  perfectly smooth at 
 $x=0,$ the point where the black hole and cosmic horizons cross.
 
Equation \ref{GRcurve} gives a detailed relation for how the entropy deficit varies with the parameter $x.$ In principle the ratio $\frac{\Delta S(x) }{\Delta S_N}$ could have been a good deal more complicated, depending in an arbitrary way on $x$ and the entropy $S.$  We will see in section \ref{dSmatrix} that the relation in \ref{GRcurve} is characteristic of theories with matrix degrees of freedom.

\section{Probabilities and the Entropy Deficit} \label{Prob and def}
The importance of the entropy deficit is that it determines probabilities for Boltzmann fluctuations through the formula \eqref{P=e-DS}

For example \eqref{S(1-xx)} implies that   the probability for the occurrence of a freak  fluctuation in which a black hole of mass $M$ appears at the pode is,
\be 
{\rm Probability}(x) = e^{-\Delta S(x) } =e^{\pi R^2 (x^2 -1)/3G}.
\label{P(x)}
\ee

The location of the black hole need not be  exactly at the pode. Let us introduce cartesian coordinates $X_i$ centered at the pode. The entropy will then depend not only on $x$ but also $X_i.$ By suitable normalization of  the coordinates the entropy deficit in \eqref{P(x)} can be generalized to  \cite{Susskind:2021omt},
\be 
{\rm Probability}(X)  = e^{-\Delta S(X) } =e^{\pi R^2(X^2 -1)/3G}.
\label{P(X)}
\ee
where $X$ represents the four component object $(x, X_i).$

Now consider the total probability for a black hole to nucleate anywhere in the static patch. It is given by an integral of the form,
\be 
{\rm Probability} = \int \frac{d^4 X}{R^4}  e^{\pi R^2 (X^2 -1)/3G}.
\ee
The range of the integration is from $X=0$ to $X \sim 1.$ The details of the
boundary at $X\sim 1$ are not important as long as the components of $X$ are of
 order $1.$ At the boundary of the integration the black hole is very small ($x \sim 1$) or its location is close to the horizon.

Defining $u=R^2 X^2$ this may be written,
\be 
{\rm Probability} = e^{-\pi R^2/3}\int_0^{R^2}  du \ u   e^{\pi u/3G}.
\label{Pint}
\ee
The integral is straightforward and gives,
\be 
{\rm Probability} = \lf \frac{3}{S_0} -\frac{9}{S_0^2}  \rg    + \lf  \frac{9}{S_0^2} \rg e^{-S_0/3}   .
\label{two terms}
\ee

 Let us rewrite \eqref{two terms} using $S_0 = \pi R^2/G,$
\be 
{\rm Probability} = \lf \frac{3G}{\pi R^2} -\frac{9G^2}{\pi^2 R^4}  \rg    + \lf  \frac{9G^2}{\pi^2 R^4} \rg e^{-\frac{\pi R^2}{3G}}   .
\label{prt  npert}
\ee
The first term in \eqref{prt  npert} appears to be perturbative in the Newton constant. It represents  contributions from very small black holes which appear close to the horizon and then fall back in. However, one might argue that this is misleading and that we should cut off the integral when the mass of the black hole becomes microscopic. In that case the first term in \eqref{prt  npert} would be replaced by something  $\exp{(-\sqrt{S})}$. This contribution  numerically dominates the second term but is  non-universal---it depends sensitively on micro-physics.

The second term, although very sub-leading,  is what really interests us. It is  non-perturbative in $G$ and due  to a saddle point in the integrand at $X=0.$ This saddle point represents the contribution of the Nariai geometry to the path integral.
 It is universal, independent of any micro-physics. 
 
One may wonder whether there is any process for which the non-universal small black hole contribution vanishes and the Nariai geometry dominates. The answer is yes; the Nariai geometry gives the  
 leading contribution to the ``inside-out" process  (see section \ref{inside out}).

\section{dS-Matrix Theory}
\label{dSmatrix}

One can argue on the basis of entropy bounds  that the holographic degrees of freedom live at the horizon of the static patch, but that argument does not tell us anything about the nature of those degrees of freedom.  I will not try to give a detailed model here, but  the  properties of de Sitter black holes can be tell us more. What we will learn is that   the degrees of freedom must be matrices  
\cite{Banks:2006rx}\cite{Susskind:2011ap}\cite{Banks:2016taq}\cite{Banks:2020zcr}.  
and the Hamiltonian should include a term whose role is to enforce certain constraints.

\subsection{Small Blocks}

Returning  to equation \eqref{DS=sqrtSs} for small black holes,
$$\Delta S = \sqrt{Ss},$$
this relation contains a hint about the nature of the holographic degrees of freedom of de Sitter space. Following Banks and collaborators \cite{Banks:2006rx} it motivates us to conjecture that the degrees of freedom are matrices (see also \cite{Susskind:2011ap}\cite{Banks:2016taq}\cite{Banks:2020zcr})  in the same sense as in M(atrix) theory  \cite{Banks:1996vh}.
To see why let's assume that the horizon degrees of freedom are a collection of $N\times N$ Hermitian matrices,
\be 
A_{m,n} = 
\begin{pmatrix}
a_{1,1} & a_{1,2} & \cdots & a_{1,N} \\
a_{2,1} & a_{2,2} & \cdots & a_{2,N} \\
\vdots  & \vdots  & \ddots & \vdots  \\
a_{m,1} & a_{m,2} & \cdots & a_{m,N} 
\end{pmatrix}
\label{mtrix}
\ee
An implicit index runs over some finite range and may include both bosonic and fermionic matrices. 
Taking a cue from BFSS M(atrix) theory \cite{Banks:1996vh} we may think of the index $m$ as running over a set of $N$ $D0$-branes. Very roughly the diagonal elements represent positions of the branes, while the off-diagonal elements represent operators which create and annihilate strings connecting the $D0$-branes.

 For now we will not specify any particular form for the Hamiltonian but we will assume that one exists, as well as  a thermal ensemble  at the appropriate  temperature. We also assume that the total entropy in thermal equilibrium is proportional to the number of degrees of freedom,
\be  
S_0 = \sigma N^2
\label{S_0=sigN2}
\ee
where $\sigma$ is the entropy per degree of freedom.

Consider a state  with a black hole of entropy 
\be
s=\sigma m^2
\label{s=sigm2}
\ee
 located at the pode.  Motivated by M(atrix) Theory we assume that the degrees of freedom split into block-diagonal form with the cosmic horizon degrees of freedom forming an $(N-m)\times (N-m)$ block, and the black hole degrees of freedom forming an $m\times m$ block. The indicies labeling the large block and small blocks  will be called $I$ and $i$ respectively. The entries in the large and small blocks are 
 $a_{IJ},$   and $a_{ij}.$  The off-diagonal elements connecting the two blocks are $a_{iJ}$ and $a_{Ij}$.

 Again, motivated by M(atrix) theory, we will assume that in a state composed of two well-separated components---in this case the small black hole and the large cosmic horizon---the  off-diagonal degrees of freedom $a_{iJ}$ and $a_{Ij}$ are constrained\footnote{Banks and Fischler have proposed that the connection between localized objects and constrained states of holographic variables is the basis for understanding locality on scales smaller than that set by the cosmological constant \cite{Banks:2016taq}\cite{Banks:2020zcr}.}  to be in their ground states, and therefore carry no entropy \cite{Banks:2006rx}\cite{Susskind:2011ap}\cite{Banks:2016taq}\cite{Banks:2020zcr}. In other words the state is constrained by $2m(N-m)$ constraints which express the condition that there are no strings connecting the $D0$-branes in the two blocks. Classically these constraints take to form,
 \be 
 a_{iJ} = a_{Ij} =0.
 \label{constraints}
 \ee

 Subject to these constraints the entropy of this state is 
\be 
S_1= \sigma (N-m)^2 +\sigma m^2
\label{S_1}
\ee

Assuming $m<<N$ and working to lowest order in $m$ the entropy deficit is,
\be 
\Delta S = 2 \sigma Nm.
\label{DSsimsNm}
\ee
Using \eqref{S_0=sigN2} and \eqref{s=sigm2} 
gives
\be 
\Delta S = 2 \sqrt{Ss}
\label{almost}
\ee
which reproduces  \eqref{DS=sqrtSs} within a factor of $2.$
I don't know of any other mechanism that will accomplish this.

The factor of two discrepancy  between \eqref{DS=sqrtSs} and \eqref{almost} may be another significant hint whose meaning we will discuss in section \ref{dynamics}.

\subsection{Remark on Higher Dimensions}

In higher dimensions things are more complicated. I will quote the $d$ dimensional generalization of \eqref{DS=sqrtSs}, valid for $s<<S_0$:
\be 
\Delta S =\lf  \frac{d-2}{2} \rg \  S_0^{\frac{1}{d-2}} \ s^{\frac{d-3}{d-2}}.
\label{magic}
\ee
This formula, derived from the $d$-dimensional \S \ solution,  can be reproduced with  matrix degrees of freedom but at a cost. It is necessary to 
allow the entropy per degree of freedom to depend on $N$ according to\footnote{A different view of the  holography of higher-dimensional dS based on multidimensional matrices
was given in \cite{Banks:2016taq}\cite{Banks:2020zcr}.},
\be 
\sigma(N) \sim \frac{1}{N^{\lf \frac{d-4}{d-3}\rg }}.
\label{sparse}
\ee  
I will leave any further dicussion of the higher dimensional generalization to future work.

\subsection{Large Blocks}
Returning to the case $d=4$ let us now consider fixed values of $s/S_0.$ The entropy deficit is proportional to the number of constraints,
\be 
\Delta S = 2\sigma  m(N-m).
\label{m(N-m)}
\ee
Following \eqref{x} and using the fact that $r_{\pm}$ are proportional to the square roots of the entropies of the horizons, we define,
\be 
x=\frac{\sqrt{s} -\sqrt{S}}{\sqrt{S_0}}.
\label{matx}
\ee
Using,
\bea
S\eq \sigma (N-m)^2 \cr \cr
s\eq \sigma m^2
\label{seqsigm^2}
\eea
we find,
\be 
x= \frac{2m}{N} -1,
\label{xofmN}
\ee
which leads to a relation  identical to \eqref{GRcurve},
\bea
\Delta S(x) \eq \Delta S_N (1-x^2), \cr \cr
\Delta S_N \eq  \frac{1}{2} S_0.
\label{DSNmatrix}
\eea
As in the earlier small-block case   \eqref{almost} the only difference between the gravitationl result and the dS-matrix result is the numerical factor. At  the Nariai point $x=0$ the entropy deficit in the matrix theory is $S_0/2$ instead of $S_0/3.$ In section \ref{dynamics} we will see that there is room in the matrix theory to decrease the numerical constant in \eqref{DSNmatrix} and bring it closer to its gravitational value of $1/3.$

There is nothing inevitable about the relation \eqref{DSNmatrix}. It does not follow from any general statistical or thermodynamic   principles. It is a consequence of the matrix degrees of freedom\ and the particular assumptions concerning the way the system is decomposed into subsystems. 
The close correspondence between the matrix theory  and general relativity calculations of $\Delta S(x)$ seems remarkable to me. I don't know any other holographic mechanism that can  lead to it.  However it is important to explain the discrepancy between the numerical factors. In the next section we will see that there is      plenty of room in the matrix theory to modify  the constants and bring them into alignment with their gravitational counterparts.

\section{Dynamics of the Constraints} \label{dynamics}

The degrees of freedom and Hamiltonian of the static patch are highly constrained by the symmetries of de Sitter space \cite{Susskind:2021omt}. Implementing those symmetries is a very hard problem which I will not try to solve in this paper.  My purpose is more modest; namely  to illustrate a dynamical mechanism  for how the constraints \eqref{constraints} can be enforced by energy considerations. 

Let us add to the matrix degrees of freedom \eqref{mtrix} one more $N \times N $ matrix denoted by$\CR,$
\be 
\CR_{m,n} = 
\begin{pmatrix}
r_{1,1} & r_{1,2} & \cdots & r_{1,N} \\
r_{2,1} & r_{2,2} & \cdots & r_{2,N} \\
\vdots  & \vdots  & \ddots & \vdots  \\
r_{m,1} & r_{m,2} & \cdots & r_{m,N}.
\end{pmatrix}
\label{rmtrix}
\ee
The notation is chosen to indicate that the eigenvalues of $\CR$ represent radial position in the static patch. 

To enforce the constraints we will assume the Matrix-theory Lagrangian contains the term,
\be 
L =   \sum  {\Tr} \lf  c^2 R^4 {\dot{A}}^2    -[\CR, A][A, \CR]              \rg
\label{Ham}
\ee
where $c$ is a numerical constant and the
sum in \eqref{Ham} is over all the other matrices---bosonic and fermionic\footnote{For fermionic matrices the quadratic kinetic term in \eqref{Ham} should be replaced by the usual Dirac term linear in time derivatives.}
---that comprise the degrees of freedom of the matrix theory.

Now consider a configuration representing an object well separated from the cosmic horizon. For simplicity the object could be at the pode at $r=0.$ The cosmic horizon is at $r=R.$ To represent this we assume the matrix $\CR$ has approximately block-diagonal form,
\bea
r_{IJ} \eq R\delta_{IJ}  +  \epsilon_{IJ} \cr 
r_{ij} \eq  \epsilon_{ij}    \cr 
r_{iJ} \eq  \epsilon_{iJ}  \cr 
r_{Ij} \eq   \epsilon_{Ij} .
\label{Req}
\eea
where $\epsilon$ is a numerically small matrix representing quantum fluctuations.

Ignoring $\epsilon,$ the commutator term  in \eqref{Ham} gives
\be 
\Tr \sum[\CR, A][A, \CR]    = R^2 \sum |a_{iJ} |^2.
\ee
Combining this with the kinetic term in \eqref{Ham} gives,
\be 
L = \sum \lf   c^2 R^4 {\dot{a}}_{iJ} {\dot{a}}_{Ji} - R^2 a_{iJ}a_{Ji} \rg
\ee
The effective Hamiltonian for the off-diagonal elements is a sum of harmonic oscillator Hamiltonians with frequency,
\be 
\omega =  \frac{1}{cR}.
\ee
If $\omega$ is much larger than the  other energy scales 
 the oscillators will be forced to their ground states and the off-diagonal degrees of freedom will carry no entropy. In that case the analysis leading up to equations \eqref{almost} and \eqref{DSNmatrix} applies unmodified.
 
 The  energy scale with which $\omega$ is to be compared is the temperature 
 $T=1/(2\pi R).$ If the numerical constant $c$ is much smaller than $2\pi$ then the constraints will be tightly enforced, but the more interesting situation is when $c\sim 2\pi.$
 In that case the constraints will not be tightly enforced; the off diagonal elements will carry some entropy, but only a fraction of $\sigma$ (the thermal entropy per degree of freedom in \eqref{S_0=sigN2}). If we carry out the analysis leading up to equations \eqref{almost} and \eqref{DSNmatrix} we will find that the only effect of relaxing  the constraints is to change the numerical coefficients in these equations. For example it should be possible to choose $c$ so as to change the constant in \eqref{almost} from $2$ to the gravitational value $1.$ At the same time that will decrease the value of factor $1/2$ in \eqref{DSNmatrix} but to bring it to exactly $1/3$ would require subtle and possibly fine-tuned properties of $H$. One might speculate that if the Hamiltonian satisfies the symmetry requirements of de Sitter space, this would be automatic.

\section{The Inside-Out Process } \label{inside out}

Classically the Nariai geometry is stable, but not quantum-mechanically \cite{Bousso:1997wi}. Initially the two horizons are at the same temperature  (see appendix \ref{Temp}) 
Now suppose a statistical fluctuation occurs and the left horizon emits a bit of energy which is absorbed by the right horizon. The effect is to increase $T_L$ and decrease $T_R.$ This creates a tendency (heat flows from hot to cold) for more energy to flow from left to right. The statistical tendency is for the left horizon to shrink down to a small black hole, while the right horizon grows to the full size of the de Sitter horizon. Eventually the small black hole will disappear, transferring all its energy to the cosmic horizon on the right side. Of course it could have happened the other way---the right horizon shrinking and the left growing.

How long does the entire process of evaporation take? The answer is roughly the Page time $t_p \sim S_0 R.$ Note that this process does not violate the second law---the entropy increases from $\frac{2}{3}S_0$ to $S_0.$

But now instead of running the system forward in time with $e^{-iHt}$ we run it backwards with $e^{iHt}.$ What will happen is the time reverse in which the system back-evolves to some micro-state of de Sitter space with either the left or right horizon growing. This implies that there are fluctuations in  the thermal state which begin with dS,  pass through Nariai space and eventually decay back to dS.
The entire history from dS to N to dS is a massive  Boltzmann fluctuation in which the de Sitter horizon emits a small black hole which then grows to the Nariai size, and then one of the two Nariai horizons shrinks back to nothing, while the other grows back to the dS size. 

\begin{figure}[H]
\begin{center}
\includegraphics[scale=.3]{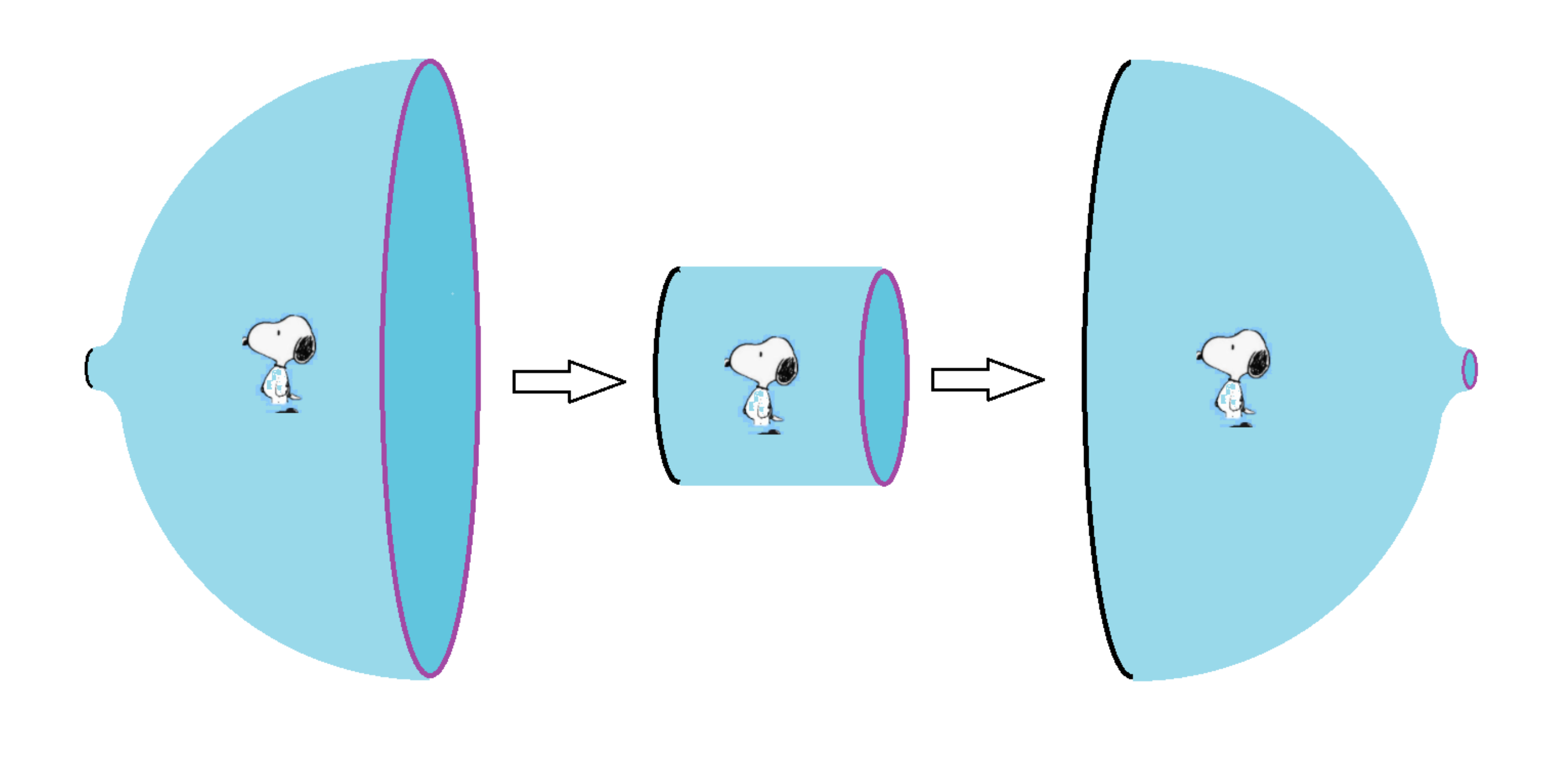}
\caption{The inside-out transition as seen by an observer in the static patch.}
\label{snoopy}
\end{center}
\end{figure}

In particular, the process can proceed so that the two horizons are exchanged. One may think of it, in terms of the diagram in figure \ref{D(x)},  as a process in which the system migrates from $x=-1$ to $x=+1,$ passing through the Nariai state at $x=0.$ This process of exchange of the horizons is the ``inside-out" process.
An observer (figure \ref{snoopy}) watching this take place would literally see the dS turn itself inside out---the tiny black hole growing and becoming the surrounding cosmic horizon while the cosmic horizon shrinks to a tiny black hole (or no black hole at all).

For the inside-out process to take place the system must pass through the Nariai state at $x=0$. Since the probability for this is $e^{-S_0/3}$ it is obviously not allowed  perturbatively.  Passing through the Nariai state gives the leading  contribution to the transition $(x=-1) \to (x=1).$ 

It is tempting to think of the inside-out process as a quantum tunneling event mediated by some kind of conventional instanton, i.e., a solution of the classical Euclidean equations of motion  interpolating from $x=-1$ to $x=+1$. This is not correct---there is no such solution. What does exist is the classical Nariai solution eternally sitting at the point $x=0.$ This is similar to a process in which a system gradually thermally up-tunnels over a broad potential barrier, mediated by an so-called  Hawking Moss instanton \cite{Hawking:1981fz}. In the Hawking-Moss framework the exponential of the Euclidean action (in this case the action of the Euclidean Nariai geometry) gives the probability to find the system at the top of the potential \cite{Weinberg:2006pc}; in other words at the Nariai point. The probability is given by
 $e^{-S_0/3}.$

\begin{figure}[H]
\begin{center}
\includegraphics[scale=.4]{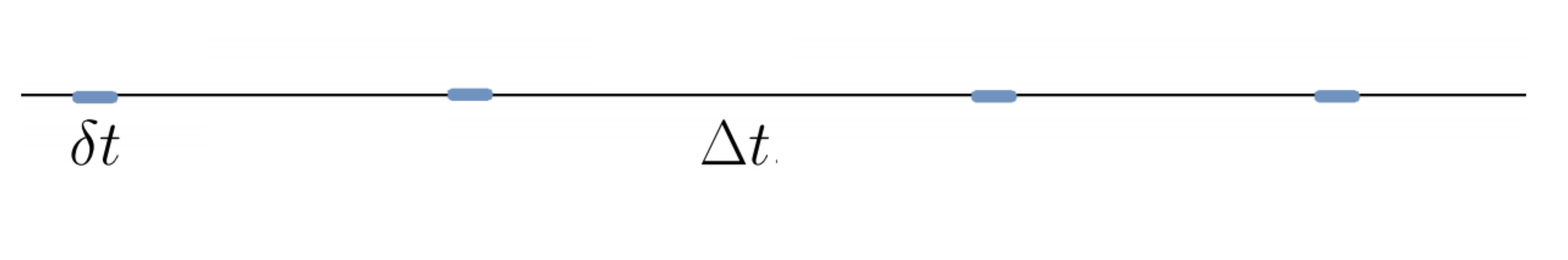}
\caption{Time scales for the inside-out process. The short blue intervals $\delta t$ represent the duration of process which takes place over a time of order the Page time. The long intervals $\Delta t$ between them represent the times between inside-out transitions.}
\label{timescales}
\end{center}
\end{figure}

The HM instanton does control the rate at which such inside-out processes occur. There are two time scales of interest. The first which I'll call $\delta t$ is how long does the process take from beginning ($x=-1$) to end  ($x=1 $)? The answer is that it takes a time of order the Page time, $\delta t \sim t_{page} \sim S_0 R.$ The other time scale, $\Delta t,$ is the average time between inside-out events. That time is very much longer: $\delta t$ is essentially instantaneous on the longer time scale
$\Delta t.$ This is shown in figure \ref{timescales}.

Under this circumstance the probability to find the system close to the Nariai state would be the ratio,
\be 
{\rm Probability} = \frac{\delta t}{\Delta t}
\ee
The probability to find the Nariai state is of order $e^{-S_0/3}$ from which we conclude,
\be 
\Delta t \sim RS_0 e^{S_0/3}.
\label{time}
\ee
The prefactor in \eqref{time} is not very reliable but it does show that the rate of inside-out events is determined by the exponential $e^{-S_0/3}.$ 
This can be compared with the longest possible decay time for  Coleman DeLuccia tunneling to a terminal vacuum, if in fact such decays are allowed. 
That time scale can in principle be as long as $e^{S_0}$ although it can be much shorter. If we suppose the decay rate to terminal vacua is as long as possible then there is plenty of time for the inside-out process to occur many times before the de Sitter vacuum decays.

The inside-out process is especially interesting because its rate is controlled by the Nariai saddle at $x=0,$ with no contribution from small black holes. In section \ref{Prob and def} the Nariai saddle was a tiny subleading effect in the probability for a black hole fluctuation, but the inside-out transition can only occur if the system passes through the Nariai point. Therefore the rate is determined by the universal saddle at $x=0.$

\bn

It is obvious what the inside-out transition means in the dS-matrix theory. The matrix representation of the unconstrained thermal equilibrium state has all $N^2$ degrees of freedom fluctuating in thermal equilibrium. The state with a small black hole is a constrained state \cite{Banks:2006rx}\cite{Susskind:2011ap}\cite{Banks:2016taq}\cite{Banks:2020zcr}
 represented by block-diagonal matrices; one small block  for the black hole, and one large block for the cosmic horizon. In the inside-out process  the small block grows while the large block shrinks until they become equal, and then continues until the blocks are exchanged. In the process the system must pass through the configuration with two equal blocks which is the matrix version of the Nariai geometry.

\section{Instantons and Giant Instantons}

The processes of small black hole formation, and the inside-out transition, exhibit some interesting parallels with instanton-mediated processes in large-N  gauge and matrix theories. 

\subsection{Some Probabilities }

This subsection summarizes the results of some probability calculations in gravity and dS-matrix theory so that we can compare them with instanton amplitudes.

The thermal formation of the smallest black hole---one with entropy of order unity---has a matrix theory description in which the small block is a single matrix element and the number of constrained is $\sim 2N$.
 The entropy deficit is 
\be 
\Delta S \sim N
\label{DSform1}
\ee
and the corresponding probability is
\be 
P = \exp{(-N)}.
\label{Pform1}
\ee

Now consider the probability in de Sitter gravity  for a minimal size black hole with $s=1,$
\be 
P = \exp{(-\sqrt{S})}.
\label{Pfors1}
\ee
Using $S=\sigma N^2$ we see that \eqref{Pform1} and \eqref{Pfors1} are essentially the same.

Next:  a  bigger fluctuation in the matrix theory, namely a fluctuation all the ways to the matrix version of the Nariai state $m=N/2$, in which the two blocks are equal. The entropy deficit is,
\be 
\Delta S = \frac{ \sigma N^2 }{2}
\label{DSNmat}
\ee
Compare that with the gravity result for the same process,
\be 
\Delta S = S_0/3.
\label{DSgrav}
\ee
Using $S=\sigma N^2$ shows that \eqref{DSNmat} and \eqref{DSgrav} scale in the same way.

\subsection{Instantons }

\begin{figure}[H]
\begin{center}
\includegraphics[scale=.4]{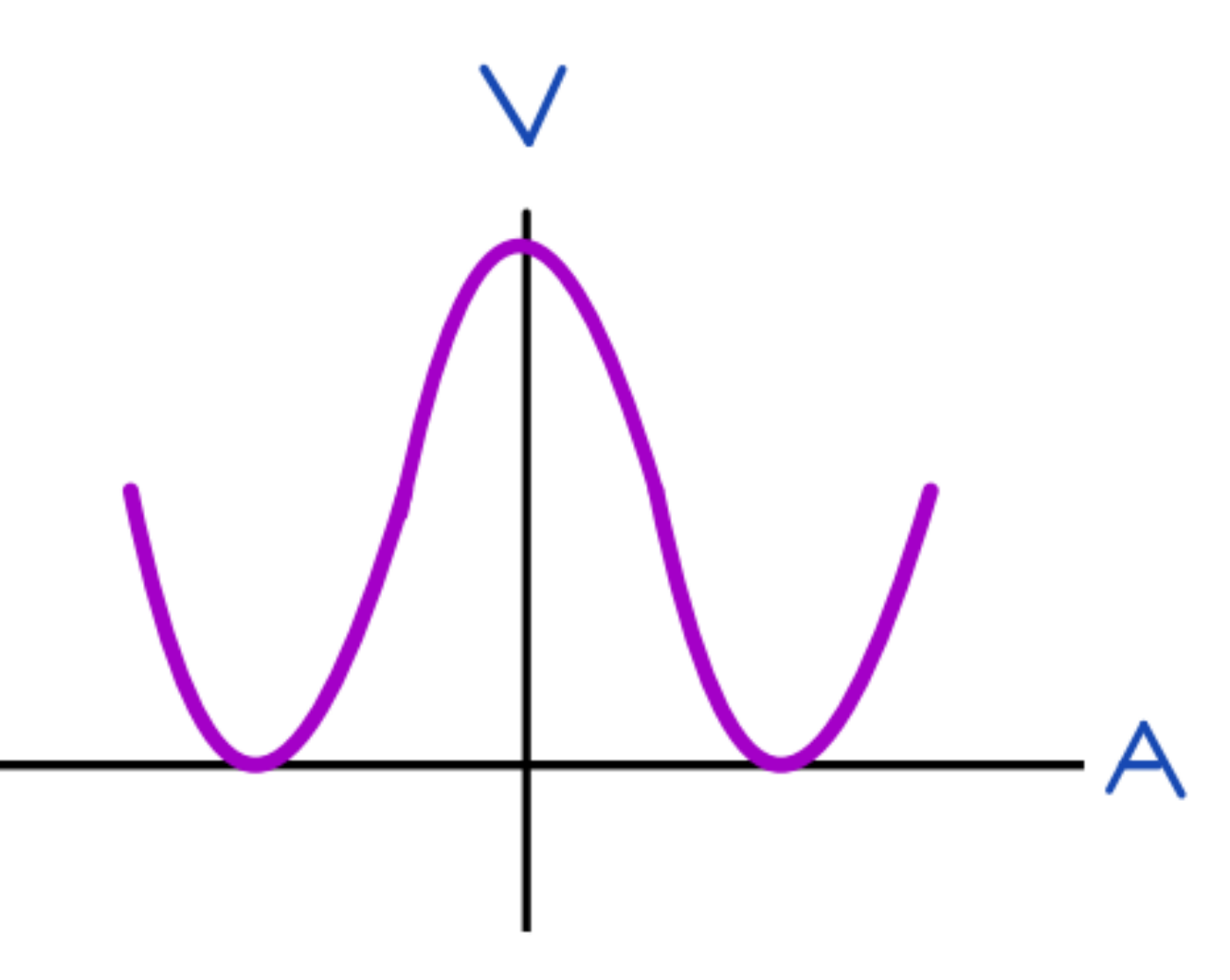}
\caption{A double-well potential for a matrix model.}
\label{VofA}
\end{center}
\end{figure}

Now let us turn to instantons, first in matrix quantum mechanics and then in gauge theories.
 The simplest example is single-matrix quantum mechanics with Lagrangian,
\be 
L = \frac{1}{2g^2} \ \lf \Tr {\dot A}^{\dag} {\dot A} -\Tr \ V(A) \rg,
\ee
and $V$ being a double well potential like the one in figure \ref{VofA}.
By standard arguments this can be reduced to the quantum mechanics of a one dimensional system of $N$ fermions which represent the eigenvalues of $A$. 

An individual eigenvalue can tunnel from the left well to the right well with probability given by an instanton. The probability for a single eigenvalue tunneling is\be  
P_1 \sim \exp{(-1/g^2)}.
\ee
In the 't Hooft large-N limit,
\be  
g^2N=\lambda \ \ \ \ \ \ \ \ \ (\lambda\sim1).
\ee
We find,
\be  
P_1 \sim \exp{(-N)}.
\ee
This simple instanton process scales with $N$ the same as in \eqref{Pform1}, suggesting that the formation of a Planck-mass black hole is an instanton-mediated process in the dS-matrix theory.

\subsection{Giant Instantons}

We may also consider a process in which all the eigenvalues tunnel from one side to the other. I'll call it a ``giant instanton." The action for a giant instanton is $N$ times larger than the simple instanton and the probability for the ``giant transition" is,
\be 
P_N \sim  \exp{(-N^2)}.    
\ee
The probability for the giant  transition scales the same way as the inside-out transition, namely $\exp{-S_0}$. We note that this transition, much like the inside-out transition takes the system between states related by a symmetry.

Instantons and giant instanton transitions also exist in Yang Mills theory. Recall that an instanton in an $SU(N)$ theory lives in an $SU(2)$ subgroup and describes a tunneling  transition of the $SU(2)$ Chern-Simons invariant by one unit. The rate also scales like $\exp{(-N)}.$ 

One can also consider a transition in which all $N/2$ commuting  $SU(2)$-subgroups tunnel. The rate for such giant instantons is $\exp{(-N^2)}.$ Thus we see a common pattern governing non-perturbative transition rates in large-N  gauge theories, and matrix theories, and  also Boltzmann fluctuations in de Sitter space.

\section{Remarks about the Holographic Principle in dS}
Semiclassically, the static patch of de Sitter space is a holographic quantum system with the degrees of freedom localized at the stretched horizon. This is reasonable in  semiclassical-gravity   but 
things are more complicated in the full non-perturbative theory. Large Boltzmann fluctuations can lead to higher topologies such as the Nariai geometry, and the horizon can break up into multiple horizons. Where, under those circumstances, do the holographic degrees of freedom reside? On the outermost or largest horizon? On the union of all of the horizons? Or is the hologram more abstract and not localized at all?

\begin{enumerate}
\item The outermost horizon? Consider a state very near the Nariai limit but with one horizon being slightly bigger than other. In this case the largest horizon has entropy slightly greater than $S_0/3.$ That is clearly not enough degrees of freedom to describe both horizons which in sum have entropy $2S/3.$  Moreover the sizes of the horizons can change with time; the largest can become the smallest and vice versa. 
\item The union of horizons? This also does not seem right. The corresponding  matrix description would be that the hologram is the union of blocks, but the Hilbert space does not factor into Hilbert spaces for the blocks. This is obvious from the fact that there are off-diagonal components $a_{Ij}$ and $a_{Ji}.$ In an approximation these elements may be unexcited if the constraints are tight, but in order to match the numerical coefficients the constraints cannot be infinitely tight. 
\item The dS-matrix theory shows that the hologram is a single  system,  with a number of degrees of freedom as large as the largest area of the cosmic horizon when it forms a single connected whole. It is large enough to describe any state of the system but not much larger. In that sense it may be identified with the horizon of the dominant saddle point in the path integral. In individual branches of the wave function no single component of the horizon may be large enough to describe the whole, but the hologram itself is.

\end{enumerate}

\section*{Acknowledgements}

I thank Adam Brown for many discussions on the material in this paper.

\bn
LS was supported in part by NSF grant PHY-1720397.

\appendix
  \section{Nariai Geometry}\label{ANG}

As the mass $M$ of the black hole increases the two solutions $r_-$ and $r_+$ come together. The limiting geometry is the Nariai solution. It occurs at the point where $dg/dr = 0.$ From \eqref{gee},  
\be 
r_N = \frac{R}{\sqrt{3}}.
\label{rN}
\ee

The entropy of each  horizon is given by $ { \frac{\rm area \it  }{4G}} = \pi r_N^2 =S_0/3.$ The combined entropy of the two horizons is the total Nariai entropy $S_N,$
\bea 
S_N &=& \frac{2}{3} S_0  \cr \cr
\eq \frac{2\pi}{3}R^2.
\label{SN}
\eea

The entropy deficit of a state $\rho$ is defined as the difference of the de Sitter entropy and the entropy of $\rho. $ For the Nariai state the entropy deficit is,
\bea
\Delta S_N \eq S_0 -S_N \cr \cr
\eq \frac{S_0}{3}. 
\label{DSN}
\eea

To understand the Nariai geometry we  begin with the near-Nariai geometry in which the two roots $r_{\pm}$ in figure \ref{cubic} are very close as can be seen from figure \ref{cubic}. The function $f(r)$ in the small interval between the roots may be expanded to quadratic order. Define,
\be 
r-\frac{R}{\sqrt{3}} =u.
\ee
Then $f$ is given by,
\be 
f= \frac{3}{R^2}(\epsilon^2 -u^2)
\label{f(e,u)}
\ee
and,
\be 
ds^2 =-  \frac{3}{R^2}(\epsilon^2 -u^2)  + \frac{R^2}{3(\epsilon^2 -u^2)}du^2 +\frac{R^2}{3} d\Omega^2.
\label{narmet1}
\ee

Now define,
\bea
v\eq \frac{u}{\epsilon} \cr \cr
\tau \eq \frac{3\epsilon}{R^2}t
\eea
and the  Nariai metric becomes,
\be 
\frac{R^2}{3} \lf  -(1-v^2) d\tau^2 + \frac{1}{1-v^2} dv^2 + d\Omega^2                                \rg.
\label{ds2xs2}
\ee
The Nariai geometry is $dS_2 \times S_2$. The Euclidean continuation is simply $S_2 \times S_2$ with both $S_2$-factors having radius $R/\sqrt{3}.$

This casts a new light on the second term of \eqref{prt  npert}. It is evidently the saddle point contribution to the path integral coming from the classical Nariai geometry---a geometry with a different topology. The contribution to the path integral can be worked out by calculating the action $I$ of the classical $S_2\times S_2$ geometry. One finds the unsurprising result
\be 
I_{Nariai} = S_N.
\label{I=S}
\ee

Consider the contributions to the Euclidean path integral from the original  de Sitter space $S_4,$ and the Nariai space $S_2\times S_2$. Schematically (ignoring prefactors), the path integral is given by,

\be 
 e^{S_0} + e^{2S_0/3} =
e^{S_0} \lf 1+e^{-S_0/3} \rg.
\label{PI}
\ee

The geometry of de  Sitter space is the non-compact goup $O(4,1)$ which is a continuation of the compact group $O(5).$ The compact group describes the symmetry of the Euclidean continuation of dS, namely the $4$-sphere. $O(4,1)$ and 
$O(5).$ are the symmetries of the semiclassical theory.
The geometry of Nariai space is $dS(2)\times S(2),$ or in the Euclidean signature,
$dS(2) \times S(2).$ The symmetry of Nariai space is $O(2,1) \times O(3) $ 
or $O(3)\times O(3).$ 
The dS symmetry is larger than the Nariai symmetry, but it does not contain the Nariai symmetry as a subgroup.

In the semiclassical limit in which the entropy is infinite the probability of transitions between the two geometries is zero. There is no obstacle to the symmetries being realized. But as we have seen, in the full quantum theory there are transitions, and it does not seem possible for either $O(4,1)$ or $O(2,1) \times O(3)$ to be exact. This clash was discussed in  \cite{Susskind:2021omt}.

\section{The Equilibrium Shell}\label{E shell}

A non-relativistic particle at rest in static coordinates will be in equilibrium at a point
 where $\frac{df(r)}{dr}=0.$ 
 For an ordinary \S  black hole in flat space no such point
 exists; $f(r)$ is monotonic in that case. 
But the \S -de Sitter 
black hole does have an equilibrium point. From \eqref{SdS} we see that the (unstable) equilibrium shell is given by,
\be 
r_{eq}^3 =MGR^2
\ee

\begin{figure}[H]
\begin{center}
\includegraphics[scale=.3]{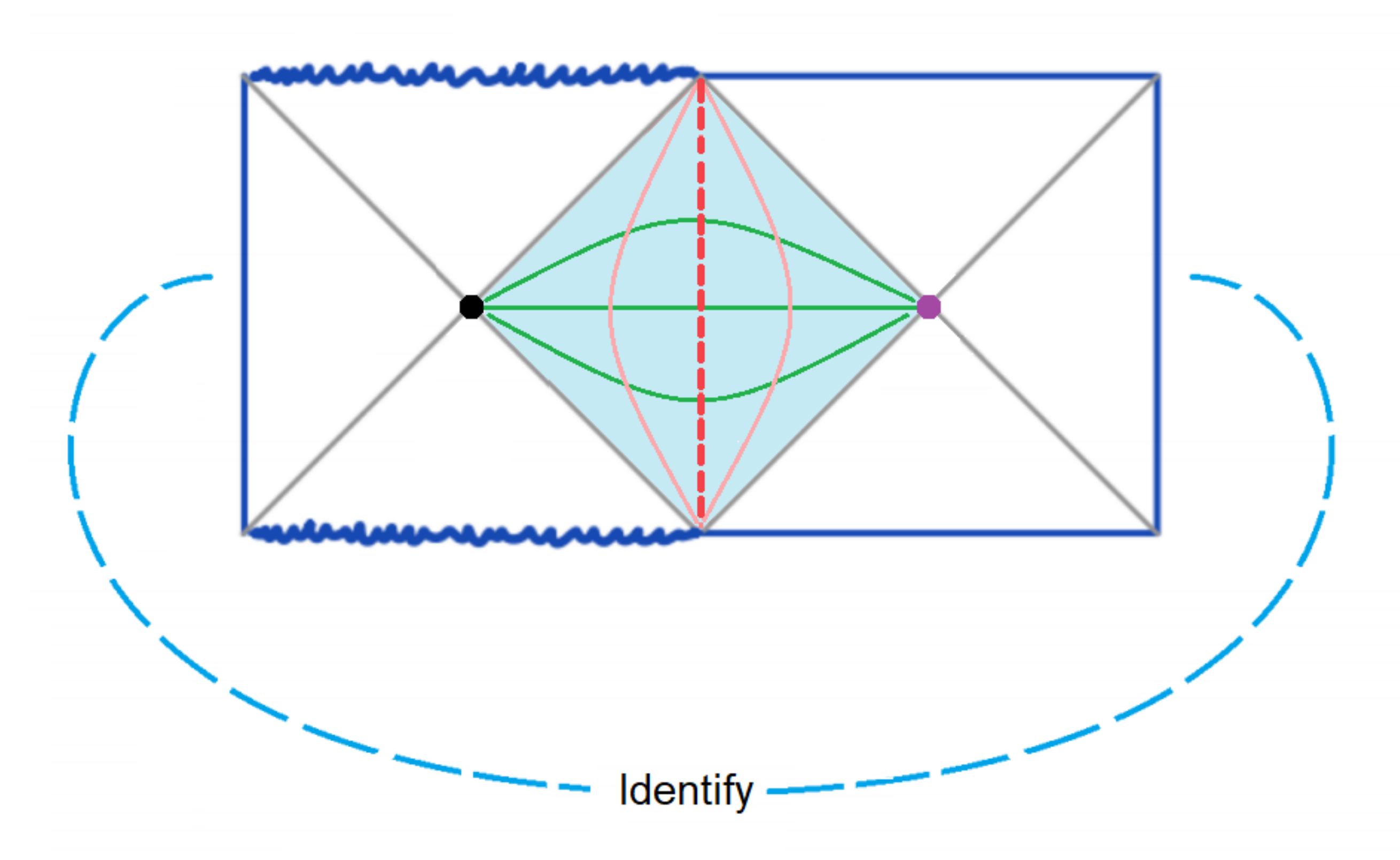}
\caption{Penrose diagram for \S -de Sitter black hole and a static patch along with the equilibrium surface shown as a red dashed line.}
\label{schds}
\end{center}
\end{figure}

Consider the Penrose diagram for a Schwarzschild-de Sitter black hole in figure \ref{schds}.

%
The static patch shown in light blue surrounds the black hole so that the black hole remains static at the center of the static patch. The dotted red line is the equilibrium shell that also surrounds the black hole at the equilibrium position.

The equilibrium shell is a natural place to introduce  observers. We can think of it as a substitute for the pode. If the observer looks in one direction he sees the black hole horizon and in the other direction he sees the cosmic horizon surrounding him. 

\begin{figure}[H]
\begin{center}
\includegraphics[scale=.5]{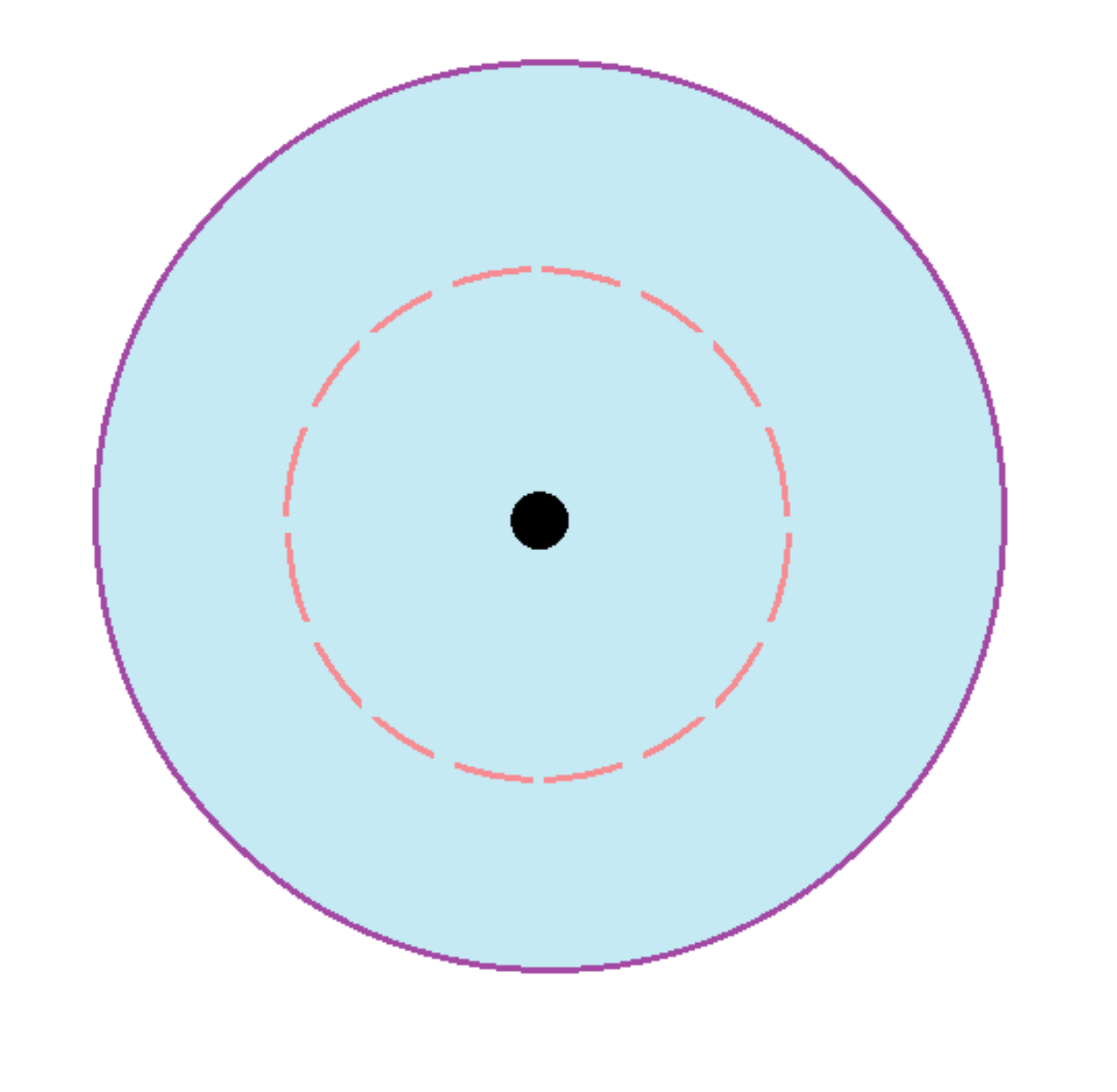}
\caption{A small black hole in the static patch. The red dashed curve is the equilibrium surface.}
\label{sds2}
\end{center}
\end{figure}

Figure \ref{sds2} shows the static patch surrounded by the cosmic horizon and the black hole at the center of the patch. The dotted red circle (really a sphere) represents the equilibrium shell.

In the Nariai limit the metric takes the form \eqref{ds2xs2} and one sees that the equilibrium shell is at the symmetry point $v=0,$ midway between the horizons. One should note that although the value of $r$ is the same at the two horizons, the distance between them is not zero. It is given by,
\bea
{\rm distance} \eq \frac{R}{\sqrt{3}}\int_{-1}^{1} \frac{1}{\sqrt{1-v^2}} dv  \cr \cr
\eq \frac{\pi R}{\sqrt{3}}.
\eea

\section{The Temperature of the Nariai Geometry}\label{Temp}

Now let us consider the Minkowski-signature Nariai geometry \eqref{ds2xs2}. In this limit the two horizons become equal and the geometry is symmetric with respect to a reflection about $v=0.$ The equilibrium shell is the two-sphere at $v=0.$ 

Let us consider the temperature of the Nariai geometry. The temperature of a black hole in flat or AdS space is usually defined as the temperature registered by a thermometer located at spatial infinity. In de Sitter space there is no asymptotic spatial infinity so we must choose another rule for defining the temperature. One possibility is to locate the thermometer at the equilibrium shell. To compute the temperature we may consider the Euclidean continuation and compute the circumference of the time-circle and identify it with inverse temperature. Alternatively we may use the fact that the Minkowski geometry is $dS_2 \times S_2$ with a radius $R/\sqrt{3}.$  It follows from both arguments that the temperature at the equilibrium shell is,
\be 
T_N=\frac{\sqrt{3}}{2\pi R}
\label{TN}
\ee
(which is larger by a factor $\sqrt{3}$ than the temperature of the original de Sitter space.)  The observer at the equilibrium position is bathed in radiation at temperature $T_N.$

Being at the same temperature, the two horizons are in thermal equilibrium with each other, but the equilibrium is unstable.

We should note that the pode of the two dimensional de Sitter space is exactly the point $v=0,$ so in that sense the temperature at the equilibrium shell is the temperature at the pode of $dS_2.$ The temperature of the original $dS_4$ is the proper temperature at the 4-D pode, and is given by,
\be 
T= \frac{1}{2\pi R.}
\label{TdS}
\ee

\end{document}